\documentclass{PoS}
\usepackage{dsfont}

\title{Bag representation for composite degrees of freedom in lattice gauge theories with fermions   }

\ShortTitle{Bag representation for composite degrees of freedom}

\author{ \speaker{Carlotta Marchis } \thanks{ This work is supported by the Austrian Science 
Fund FWF, grant I 2886-N27 and the FWF DK W 1203, ''Hadrons in Vacuum, Nuclei and Stars".}        \\
        University of Graz, Institute of Physics, 8010 Graz, Austria\\
        E-mail: \email{carlotta.marchis@uni-graz.at}}
        
\author{Christof Gattringer\\
        University of Graz, Institute of Physics, 8010 Graz, Austria\\
        E-mail: \email{christof.gattringer@uni-graz.at}}
       
\author{Oliver Orasch\\
        University of Graz, Institute of Physics, 8010 Graz, Austria\\
        E-mail: \email{oliver.orasch@uni-graz.at}}

\abstract{We explore new representations for lattice gauge theories with fermions, where the space-time lattice is divided into dynamically fluctuating regions, inside which different types of degrees of freedom are used in the path integral. The first kind of regions is a union of so-called bags, in which the dynamics is described by the free propagation of composite degrees of freedom of the original fermions. In the second region, called complementary domain, configurations of the remaining interacting degrees of freedom are used to describe the dynamics. We work
out the bag representation for the gauge groups SU(2) and SU(3) and address the nature of the strong coupling 
effective degrees of freedom, which are fermions for SU(3) and bosons for SU(2). We discuss first steps towards a 
numerical simulation of the bag representations.}

\FullConference{The 36th Annual International Symposium on Lattice Field Theory - LATTICE2018\\
		22-28 July, 2018\\
		Michigan State University, East Lansing, Michigan, USA.}

\begin{document}

\section{Introduction}
\noindent
In recent years an important development in lattice field theories was the construction of 
worldline and worldsheet representations (see, e.g., the reviews at the annual lattice conferences 
\cite{reviews}). Often the motivation are lattice field theories at finite density where in some
cases the complex action problem may be overcome by a suitable worldline/worldsheet representation. 
For systems with fermions, such as QCD, the situation is more complicated due to the Grassmann nature of the 
fermions and additional signs from the Clifford algebra. However, for strong coupling worldline
representations have been studied since the early days of lattice field theory \cite{classic}. 
This idea has been revisited recently and interesting results were obtained for strong coupling QCD 
\cite{recent,Gattringer:2018mrg}. Also beyond strong coupling suggestions for a fully 
dualized version of lattice QCD \cite{weak} were presented.

Also for purely fermionic lattice field theories worldline representations play an important role. In some cases it 
is possible to develop a representation in terms of so-called fermion bags. Fermion bags are space-time 
domains on the lattice where the dynamics is essentially given by free fermions which inside the bag can be described
by a fermion determinant. In between the bags the fermionic Grassmann integral is saturated 
with the interaction terms \cite{bags1}. For many interesting systems fermion bag representations
were applied as a powerful tool for numerical simulations 
\cite{bags2}. 

In this contribution we discuss further developments based on \cite{Gattringer:2018mrg}, 
where a so-called baryon bag 
representation was constructed for strong coupling lattice QCD. Although the technical aspects of a simulation 
with baryon bags are similar to the fermion bag concept, the underlying physical picture is different: Inside a
baryon bag three strongly coupled quarks propagate jointly as a baryon which is described as a free composite 
fermion (whereas inside a conventional fermion bag the fundamental fermions of the theory propagate as
free fermions). The physics inside the baryon bag is described by a free fermion determinant, while outside the 
baryon bags the dynamics is governed by quark and diquark monomer and dimer terms.  
Here we report on developments towards an efficient simulation of the monomers and dimers and also 
discuss the case of strong coupling SU(2), where the degrees of freedom inside the bags are 
nilpotent bosons, that are described by bag permanents instead of the bag determinants that appear for fermionic 
effective degrees of freedom. 

\section{Factorization of strong coupling QCD}
\noindent
We begin our presentation with a brief summary of the baryon bag formulation \cite{Gattringer:2018mrg} 
of strong coupling QCD. We use one flavor of staggered fermions, where the action is given by 
\begin{equation}
S_{F}\big[\overline{\psi}, \psi, U\big]  \; = \; \sum_{x} \biggl( 2m \, \overline{\psi}_x \psi_x \; +
\; \sum_{\nu} \gamma_{x,\nu} \Big[ \,  \overline{\psi}_x U_{x,\nu} \psi_{x + \hat{\nu}}  - 
 \overline{\psi}_{x + \hat{\nu}} U_{x,\nu}^{\dagger} \psi_x \Big] \!\biggr) \; .
\label{quarkaction}
\end{equation}
$\psi_x$ and $\overline{\psi}_x$ are Grassmann variables with 3 color components and by $\gamma_{x,\nu}$ 
we denote the staggered sign factors $\gamma_{x,1} = 1$, $\gamma_{x,2} = (-1)^{x_1}$, $\gamma_{x,3} = (-1)^{x_1+x_2}$ and $\gamma_{x,4} = (-1)^{x_1+x_2+x_3}$. The gauge fields are coupled via the link variables
$U_{x,\nu} \in$ SU(3). At strong coupling the partition sum is given by
\begin{equation}
Z \; = \; \int \!\! D [\, \overline{\psi}, \psi ] \int \!\! D[U]\, e^{\,S_F\big[\overline{\psi}, \psi, U\big]}
\; = \;  \int \!\! D [\, \overline{\psi}, \psi ] \prod_x \!e^{2m \, \overline{\psi}_x \psi_x } \!\!\! \int \!\! D[U]\, 
\prod_{x,\nu}  e^{\, \gamma_{x,\nu} \, \overline{\psi}_x U_{x,\nu} \psi_{x + \hat{\nu}} } 
e^{\, -\gamma_{x,\nu} \, \overline{\psi}_{x + \hat{\nu}} U_{x,\nu}^{\dagger} \psi_x } ,
\end{equation}
where in the second step we have written the sums in the action as products of the individual Boltzmann factors.
The exponentials of the nearest neighbor terms can be expanded into power series that terminate after the third 
order, since for gauge group SU(3) we have only 3 independent (pairs of) Grassmann variables per site (for notational 
convenience we here drop the space-time indices and use matrix-vector notation for the color indices)
\begin{equation}
e^{\, \gamma \, \overline{\psi} \, U \psi } \; = \;  
1  + \gamma \, \overline{\psi} U \psi \,  + \frac{(\overline{\psi} U \psi)^2}{2 !}  
 + \frac{(\gamma \, \overline{\psi} U \psi)^3}{3 !}  \; = \; 
\bigg[ 1 + \gamma \frac{(\overline{\psi} U \psi)^3}{3 !} \bigg] 
\bigg[1  + \gamma \, \overline{\psi} U \psi \, + \frac{(\overline{\psi} U \psi)^2}{2 !}  \bigg] \; .
\label{aux0}
\end{equation}
In the second step we have pulled out the highest monomial and the product of the two square brackets reproduces
the previous expression due to nilpotency. It is easy to see that the 3rd order term is
independent of the gauge link $U$:  $(\overline{\psi} U \psi)^3  =  (\overline{\psi}_a U_{ab} \psi_b)^3 = 
3! \; \overline{\psi}_3 \, \overline{\psi}_{2} \, \overline{\psi}_{1} \, \psi_1 \,  \psi_{2} \, \psi_{3} \; \det U \equiv  
3! \; \overline{B} \, B$, where in the last step we have used $\det U = 1$ and introduced the baryon fields 
$B_x$, $\overline{B}_x$ defined as
$B_x =  \psi_{x,1} \, \psi_{x,2} \, \psi_{x,3}, \; 
\overline{B}_x =  \overline{\psi}_{x,3} \, \overline{\psi}_{x,2} \, \overline{\psi}_{x,1}$. Thus we obtain
\begin{equation}
e^{\, \gamma \, \overline{\psi} \, U \psi } \, = \, e^{\gamma \overline{B} B} \;
\sum_{d=0}^2 \! \frac{ \!\! \Big( \! \gamma \, \overline{\psi} U \psi \! \Big)^{\! d}\!\!}{d!} , \;
e^{\, - \gamma \, \overline{\psi} \, U^\dagger  \psi } \, = \, e^{-\gamma \overline{B} B} \;
\sum_{\overline{d}=0}^2 \! 
\frac{\!\! \Big( \! \!- \gamma \, \overline{\psi} U^\dagger \psi \! \Big)^{\!\overline{d}}\!\!}{\overline{d}!} , \;
e^{\, 2m \overline{\psi} \psi } \, = \, e^{(2m)^3 \overline{B} B} \;
\sum_{s=0}^2 \! \frac{ \!\! \Big( \! 2m \overline{\psi} \psi \! \Big)^{\!s}\!\!}{s!} ,
\label{aux}
\end{equation}
where we have repeated the steps for the forward hopping term also for the backward hopping and mass terms.
We stress again, that all contributions that we have expressed in terms of the baryon fields $B_x$ and 
$\overline{B}_x$ are independent of the gauge fields. Using the results (\ref{aux}) we find for the partition sum 
$Z  = \int \!\! D [\, \overline{\psi}, \psi ] \, e^{ \, S_B[\, \overline{B},B]} \; 
W [\, \overline{\psi}, \psi]$.
We have factorized the contributions of the baryons and organized them in the baryon action 
(we re-inserted all space-time indices)
\begin{equation}
 S_{B}\big[ \, \overline{B}, B \big]  \; =  \; \sum_{x} \biggl( 2M \, \overline{B}_x B_x \; +
\; \sum_{\nu} \gamma_{x,\nu} \Big[ \, \overline{B}_x \, B_{x + \hat{\nu}} \, -  \, 
\overline{B}_{x + \hat{\nu}}  B_x \Big] \!\biggr) \; ,
\label{baryonaction}
\end{equation}
which has the form of a free staggered action for the baryon fields $B_x$ and $\overline{B}_x$
with mass $M = 4 m^3$. 

\section{Strong coupling integrals}
\noindent
The remaining, non-baryonic terms depend on the gauge fields and we have collected them in
\begin{equation}
 W [\, \overline{\psi}, \psi] \, = \, \prod_{x} \! \sum_{s_x} \!\!
\frac{ \!\! \Big(\! 2m  \overline{\psi}_x \psi_x \! \Big)^{\! s_x}\!\!\!\!}{s_x !} 
 \prod_{x,\nu} \sum_{d_{x,\nu}, \overline{d}_{x,\nu}} \!\!\!\!\!\!
\frac{(\gamma_{x,\nu})^{d_{x,\nu}+\overline{d}_{x,\nu}} \!\!}{d_{x,\nu}! \; \overline{d}_{x,\nu}!} \!\!
\int \!\! D[U] 
\Big(\! \overline{\psi}_x  U_{x,\nu} \psi_{x + \hat{\nu}} \! \Big)^{\! d_{x,\nu}} \!
\Big(\!\! - \overline{\psi}_{x  +  \hat{\nu}}  U_{x,\nu}^\dagger  \psi_x \! \Big)^{\! \overline{d}_{x,\nu}} \!\!\!.
\end{equation}
The integrals over the gauge links project to color singlet combinations. These are obtained either by
the trivial choice $d_{x,\nu} = \overline{d}_{x,\nu} = 0$, the quark dimer term $d_{x,\nu} = \overline{d}_{x,\nu} = 1$ 
or the diaquark dimer term $d_{x,\nu} = \overline{d}_{x,\nu} = 2$. The corresponding integrals are well known, 
see, e.g., \cite{creutz}, and we find \cite{Gattringer:2018mrg} 
\begin{equation}
Z \; = \; \int \!\! D [\, \overline{\psi}, \psi ] \, e^{ \; S_B[\, \overline{B},B]} \;
\prod_{x} \sum_{s_x = 0}^2 \frac{ \Big(\! 2m \,  \overline{\psi}_x \psi_x \! \Big)^{s_x}\!\!}{s_x !} \;
\prod_{x,\nu}  \sum_{d_{x,\nu} =0}^2 \!\!\!  \frac{(3-d_{x,\nu})!}{6 \,  d_{x,\nu}!} \big( \overline{\psi}_x  \psi_x \;\;
\overline{\psi}_{x  +  \hat{\nu}} \psi_{x + \hat{\nu}} \big)^{d_{x,\nu}} \! .
\label{zstrong1}
\end{equation}
The final step is to saturate the Grassmann integral. The first possibility is to saturate the Grassmann integral 
using the baryon terms. Since the baryon fields $B_x$ and $\overline{B}_x$ contain all three colors, the baryon 
terms completely saturate the Grassmann integral on the sites where we use them.  
The fact that there are no remaining quark link terms at strong coupling also implies that the 
regions where we saturate the Grassmann integral with the baryon terms do not mix with the other sites where we 
use the terms from $W[\, \overline{\psi}, \psi]$. We now refer to a connected area of space-time where we 
use the baryon terms for saturating the Grassmann integral as a baryon bag ${\cal B}_i$. By ${\cal B}$ 
we denote the union of 
all bags. Since the $B_x$ and $\overline{B}_x$ inherit the Grassmann properties of the 
underlying quark fields \cite{Gattringer:2018mrg}, the contribution of the baryons inside 
the bag ${\cal B}_i$ is given by the determinant $\det D^{(i)}$ of the free staggered Dirac operator $D^{(i)}$ 
from the baryon action (\ref{baryonaction}), but restricted to the sites of ${\cal B}_i$.

In the complementary domain $\overline{{\cal B}}$ we saturate the Grassmann integral by combining quark 
monomer ($s_x = 1$) and diquark monomer terms ($s_x = 2$) on a single site, with quark dimers ($d_{x,\nu} = 1$) and 
diquark dimers ($d_{x,\nu} = 2$) on links. Note that since the Grassmann variables are already in the correct
order no negative signs remain. We summarize the baryon bag form as 
\cite{Gattringer:2018mrg},
\begin{equation}
Z \; =  \; \sum_{\{{\cal B}\}} \; \! \prod_i \det \, D^{(i)}  \! \sum_{\{s,d\, \| \, \overline{{\cal B}}\}}  \!
(2m)^{\sum_x s_x} \left( \frac{1}{3} \right)^{\sum_{x ,\nu} \big[ \delta_{d_{x,\nu},1} + \delta_{d_{x,\nu},2}\big] } \; .
\label{Zfinal}
\end{equation}
The sum runs over all possible ways of decomposing the lattice into baryon bags ${\cal B}_i$ and a complementary
domain ${\overline{\cal B}}$. For each bag ${\cal B}_i$ we collect as a factor the determinant $\det D^{(i)}$ of 
the Dirac operator $D^{(i)}$ in (\ref{baryonaction}) restricted to ${\cal B}_i$. In the complementary domain we saturate 
the Grassmann integral with with quark monomers $(s_x =1)$, diquark monomers $(s_x = 2)$, 
quark dimers $(d_{x,\nu} = 1)$ and
diquark dimers $(d_{x,\nu} = 2)$, and by the second sum we denote the sum over all admissible monomer and dimer 
configurations compatible with a given complementary domain ${\overline{\cal B}}$. Quark and diquark monomers
come with a factor $(2m)^{s_x}$, quark and diquark dimers both with a factor of 1/3.

\section{Towards an efficient update strategy}
\begin{figure}[t!]
	\centering      
	\includegraphics[height=16mm,clip]{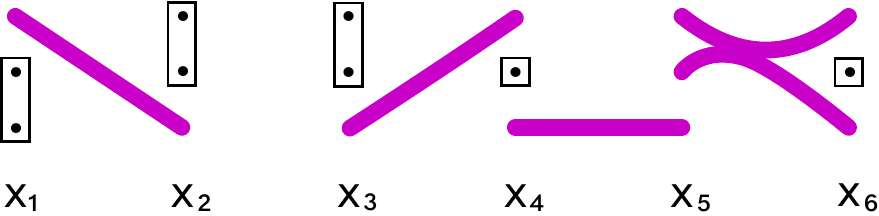}
		\vspace*{-2mm}
	\caption{Examples for saturation in the complementary domain.}
	\label{example}
	\vspace*{-3mm}
\end{figure}

\noindent
We begin the discussion of possible update strategies for a numerical simulation of the baryon bag representation 
with the complementary domain ${\overline{\cal B}}$. Here we have to occupy each site of the lattice with 
quark and diquark monomers or with the endpoints of quark and diquark dimers. More specifically each site of the 
lattice has three possible colors where we can attach the corresponding elements. In order to illustrate this, in 
Fig.~\ref{example} we show a chain of neighboring sites $x_1, \, x_2\; .... \;x_6$. For every site we use 
three layers that correspond to the three colors. A quark monomer is represented by a single square and a diquark 
monomer by a rectangle that covers two layers. Both, quark and diquark monomers live on a single site and 
we placed diquark monomers at the sites $x_1$, $x_2$ and $x_3$, while quark monomers are placed at $x_4$ 
and $x_6$. Dimer terms live on the links and we represent a quark dimer by a single fat line, while diquark dimers
are represented by two touching lines on the same link. On the links between $x_1$ and $x_2$,  
$x_3$ and $x_4$, as well as $x_4$ and $x_5$ we have placed quark dimers, while on the link between
$x_5$ and $x_6$ we have placed a diquark dimer. Obviously, in our example at every site we have all three color 
components saturated either by monomer terms or endpoints of dimer terms. We stress that in Fig.~\ref{example} 
the sites simply form a chain, but of course at some site $x$ on a $d$-dimensional lattice one has to take into 
account all $2d$ links attached to that site.

In the partition function (\ref{Zfinal}) we have to sum over all monomer and dimer assignments 
that are compatible with a given structure of the complementary domain ${\overline{\cal B}}$. We will base this 
summation on our monomer variables $s_x = 0,1,2$ on the sites and the dimer variables 
$d_{x,\nu} = 0,1,2$ based on the links. Since at each site $x$ every color has to be occupied by either a monomer
or the endpoint of a dimer term, we find the following simple constraint, 
\begin{equation}
s_x \; + \;  \sum_{\nu} [ d_{x,\nu} \, + \, d_{x-\hat \nu,\nu} ] \; = \; 3 \quad \forall \; x \in \overline{\cal B}.
\label{constraints}
\end{equation}
It turns out that for a given set of fixed monomer variables $s_x$ and dimer variables $d_{x,\nu}$ one
can sum over the multiplicities for the different configurations of monomer and dimer arrangements that 
are compatible with the fixed values of the $s_x$ and $d_{x,\nu}$. We discuss this for the examples in 
Fig.~\ref{example}: At site $x_1$ we can choose to place the endpoint of the quark dimer at any of the 
three colors and the remaining two colors are saturated with the diquark monomer. Thus at the site $x_1$
we find a degeneracy factor of $F(x_1) = 3$. The same analysis holds for the sites $x_2$ and $x_3$ that 
both have a degeneracy factor of $F(x_2) = F(x_3) = 3$. The situation is different for $x_4$: There we have 
3 choices for placing the quark monomer and another 2 choices for placing the endpoints of the two quark dimers,
giving a degeneracy of $F(x_4) = 6$. For $x_5$ and $x_6$ we again find a degeneracy of 
$F(x_5) = F(x_6) = 3$, because we have 3 choices for placing the endpoint of the quark dimer at $x_5$,
respectively the quark monomer at $x_6$ and the remaining two colors are saturated by the two endpoints 
of the diquark dimer. Thus whenever a diquark element is attached at a site $x$, the multiplicity is 3, while
it is 6 when only quark elements are used for the saturation at $x$. Thus we obtain 
$F(x) = 6 - 3 \delta_{s_x,2} - 3 \sum_{\nu}  \big[ \delta_{d_{x,\nu},2} \, + \, \delta_{d_{x-\hat\nu,\nu},2} \big]$ and
rewrite the partition sum (\ref{Zfinal}) into the form
\begin{equation}
Z \; =  \sum_{\{{\cal B}\}} \; \! \prod_i \det \, D^{(i)}  \!\!\!\!\! \sum_{\{s,d \in \overline{{\cal B}}\}} \!
\left[ \, \prod_{x \in \overline{\cal B}} \!\!
\delta\Big( s_x + \sum_{\nu} [ d_{x,\nu} \, + \, d_{x-\hat\nu,\nu} ] - 3 \Big) \;  F(x) \right]\!
(2m)^{\sum_x s_x} \left( \frac{1}{3} \right)^{\!\!\sum_{x ,\nu} \! \big[ \delta_{d_{x,\nu},1} + \delta_{d_{x,\nu},2}\big]} \!\!.
\label{Zfinal2}
\end{equation}
The first sum is again the sum over all bag configurations and we collect the factors from the corresponding 
bag determinants. By the second sum $\sum_{\{s,d \in \overline{{\cal B}}\}} \equiv
\prod_{x \in \overline{{\cal B}}} \sum_{s_x = 0}^2 
\prod_{(x,\nu) \in \overline{{\cal B}}} \sum_{d_{x,\nu} = 0}^2$ we denote the sum over all monomer and dimer 
configurations in the complementary domain. These are subject to the constraints (\ref{constraints}) at all sites $x$ 
which we write as a product of Kronecker deltas, here denoted as $\delta(n) \equiv \delta_{n,0}$. The admissible
configurations come with degeneracy factors $F(x)$ at all sites $x$ in the complementary domain, and the
weights from the quark mass and the factors 1/3 for quark and diquark dimers. In the form (\ref{Zfinal2})
the baryon bag representation is better accessible to numerical simulations and we have started to implement the
first numerical tests. 

\section{The case of strong coupling SU(2)}

\noindent
We conclude our discussion of bag representations for effective strong coupling degrees of freedom with the case of 
gauge group SU(2). In this case the effective degrees of freedom are diquarks which are bosons. Thus we expect 
a different form of the contributions from the bags: 1) The action for the bag terms should be bosonic in nature, i.e.,
we expect to find a Helmholtz operator for the kernel. 2) The bag determinant, which is fully anti-symmetric,
should be replaced by a completely symmetric expression. Since the diquarks are nilpotent, we will see that the
bag determinant is replaced by a so-called permanent which like the determinant is a sum over all permutations, 
but the sign of the permutations is omitted. 

We start our discussion of SU(2) with the expansion of the nearest neighbor Boltzmann factors as in 
(\ref{aux0}) and (\ref{aux}). Since for SU(2) we only have two colors, the power series for the Boltzmann 
factor terminates after the quadratic term. As before, this highest order term is independent of the gauge link,
$(\overline{\psi} U \psi)^2  =  
2! \; \overline{\psi}_{2} \, \overline{\psi}_{1} \, \psi_1 \,  \psi_{2} \; \det U \equiv  
2! \; \overline{B} \, B$, where now the baryonic degrees of freedom are diquarks, i.e., 
$B_x =  \psi_{x,1} \, \psi_{x,2}, \, \overline{B}_x =  \overline{\psi}_{x,2} \, \overline{\psi}_{x,1}$. We find
for the nearest neighbor Boltzmann factor
\begin{equation}
e^{\; \gamma_{x,\nu}  \, \overline{\psi}_x \, U_{x,\nu} \psi_{x + \hat{\nu} }} \; = \;  
e^{\, \overline{B}_x \, B_{x + \hat{\nu}} } \, \sum_{d_{x,\nu} = 0}^1 
\gamma_{x,\nu} \, \big(\overline{\psi}_x U_{x,\nu} \psi_{x + \hat{\nu}} \big)^{d_{x,\nu}} 
\; .
\label{aux3}
\end{equation}
In the exponential on the rhs.\ the staggered sign $\gamma_{x,\nu}$ has disappeared 
since it is squared for diquarks. The baryonic/diquark fields $B_x$ and $\overline{B}_x$ are now
made of an even number of Grassmann variables, such that they commute. Combining the
nearest neighbor terms (\ref{aux3}) with the corresponding backward hopping and the mass term we find
a bosonic action for the diquark fields 
\begin{equation}
S_{B}\big[ \, \overline{B}, B \big]  \; =  \; \sum_{x} \biggl( 4m^2 \, \overline{B}_x B_x \; +
\; \sum_{\nu} \Big[ \, \overline{B}_x \, B_{x + \hat{\nu}} \, +  \, 
\overline{B}_{x + \hat{\nu}}  B_x \Big] \!\biggr) \; = \; \sum_{x,y}  \overline{B}_x \, H_{x,y} \, B_x  \; .
\label{diquarkaction}
\end{equation}
In the last step we have defined the kernel $H_{x,y}$ of the action, which is the sum of
a constant term and the discretized Laplace operator, i.e., a discretized Helmholtz operator. Again we write the
partition sum as $Z  = \int \!\! D [\, \overline{\psi}, \psi ] \, e^{ \, S_B[\, \overline{B},B]} \; 
W [\, \overline{\psi}, \psi]$, where the weight for the non-baryonic terms is
\begin{equation}
 W [\, \overline{\psi}, \psi] \; = \; \prod_{x} \sum_{s_x=0}^1 \,
 \Big(\! 2m  \overline{\psi}_x \psi_x \! \Big)^{\! s_x} \;
\prod_{x,\nu} \sum_{d_{x,\mu}=0}^1 \; \int \!\! D[U] \,
\Big(\! -  \overline{\psi}_x  U_{x,\nu} \psi_{x + \hat{\nu}} \,
\overline{\psi}_{x  +  \hat{\nu}}  U_{x,\nu}^\dagger  \psi_x \! \Big)^{\! d_{x,\nu}} \; .
\end{equation}
We have already used the fact that only the same powers of $U_{x,\nu}$ and 
$U_{x,\nu}^\dagger$ give rise to a singlet and thus a non-trivial strong coupling integral. As a consequence 
we have the 
following non-trivial contributions from the non-baryonic terms: Quark monomers for $s_x = 1$ and quark dimers
for $d_{x,\nu} = 1$ (here the corresponding SU(2) integral \cite{creutz} gives a factor of 1/2). 

Again we can decompose the lattice into bags ${\cal B}_i$ where we use the baryonic/diquark terms to saturate the 
Grassmann integral and a complementary domain where we saturate the Grassmann integral with quark monomers
and dimers. Note that for SU(2) no diquark terms appear in the complimentary domain -- they are the degrees
of freedom used inside the bags. Before we can write down the final expression for the partition sum we need 
to clarify the form of the contribution inside the bags. We already remarked, that for strong coupling SU(2) the
baryon/diquark fields $B_x$ and $\overline{B}_x$ commute since they are composed of pairs of Grassmann 
variables. However, they are still nilpotent, such that we obtain the same algebraic form for the integral 
over the degrees of freedom in the bag, but without the signs that were generated when commuting 
the SU(3) strong coupling fields. Thus instead of the bag determinant we obtain a so-called permament
perm $H^{(i)}$, where $H^{(i)}$ is the Helmholtz operator defined in (\ref{diquarkaction}) but restricted to the 
bag ${\cal B}_i$. The definition of the permanent perm $M$ of a $N \times N$ matrix $M$ reads 
perm $M = \sum_{\pi(N)} \, \prod_{k=1}^N \, M_{k,\pi_k}$ where the sum runs over all permutations of $N$ numbers. 
The final result for the bag representation for the strong coupling SU(2) case thus reads (the degeneracy factor is 
always $F(x) = 2 \; \forall x \in \overline{\cal B}$)
\begin{equation}
Z \; =  \sum_{\{{\cal B}\}} \; \! \, \prod_i \mbox{perm} \; H^{(i)}  \!\!\!\!\! \sum_{\{s,d \in \overline{{\cal B}}\}} \!
\left[ \, \prod_{x \in \overline{\cal B}} \!\!
\delta\Big( s_x + \sum_{\nu} [ d_{x,\nu} \, + \, d_{x-\hat\nu,\nu} ] - 2 \Big) \;  F(x) \right]\!
(2m)^{\sum_x s_x} \left( \frac{1}{2} \right)^{\!\!\sum_{x ,\nu} d_{x,\nu} } .
\end{equation}
This example of a case with effective bosonic bag degrees of freedom concludes our presentation of bag techniques 
for  strong coupling gauge fields with fermions.

\end{document}